# Bound Analysis of Imperative Programs with the Size-change Abstraction (extended version)

Florian Zuleger[1], Sumit Gulwani[2], Moritz Sinn[1], and Helmut Veith[1]*

[1] TU Wien, {zuleger,sinn,veith}@forsyte.at
[2] Microsoft Research, sumitg@microsoft.com

**Abstract.** The size-change abstraction (SCA) is an important program abstraction for termination analysis, which has been successfully implemented in many tools for functional and logic programs. In this paper, we demonstrate that SCA is also a highly effective abstract domain for the *bound analysis* of *imperative programs*.

We have implemented a bound analysis tool based on SCA for imperative programs. We abstract programs in a pathwise and context dependent manner, which enables our tool to analyze real-world programs effectively. Our work shows that SCA captures many of the essential ideas of previous termination and bound analysis and goes beyond in a conceptually simpler framework.

## 1 Introduction

Computing symbolic bounds for the resource consumption of imperative programs is an active area of research [17,15,14,13,12,10]. Most questions about resource bounds can be reduced to counting the number of visits to a certain program location [17]. Our research is motivated by the following technical challenges:

**(A)** Bounds are often *complex non-linear arithmetic expressions* built from $+, *, \max$ etc. Therefore, abstract domains based on linear invariants (e.g. intervals, octagons, polyhedra) are not directly applicable for bound computation.

**(B)** The proof of a given bound often requires *disjunctive invariants* that can express loop exit conditions, phases, and flags which affect program behavior. Although recent research made progress on computing disjunctive invariants [17,14,27,8,5,29,11], this is still a research challenge. (Note that the domains mentioned in (A) are conjunctive.)

**(C)** It is *difficult to predict a bound in terms of a template* with parameters because the search space for suitable bounds is huge. Moreover the search space cannot be reduced by compositional reasoning because bounds are global program properties.

**(D)** It is not clear how to *exploit the loop structure* to achieve compositionality in the analysis for bound computation. This is in contrast to automatic termination analysis where the cutpoint technique [8,5] is used standardly.

* Research supported by the FWF research network RiSE, the WWTF grant PROSEED and Microsoft Research through a PhD Scholarship.

In this paper we demonstrate that the size-change abstraction (SCA) by Lee et al. [24,4] is the right abstract domain to address these challenges. SCA is a predicate abstraction domain that consists of (in)equality constraints between integer-valued variables and boolean combinations thereof in disjunctive normal form (DNF).

SCA is well-known to be an attractive abstract domain: First, SCA is rich enough to capture the progress of many real-life programs. It has been successfully employed for automatic termination proofs of recursive functions in functional and declarative languages, and is implemented in widely used systems such as ACL2, Isabelle etc. [25,20]. Second, SCA is simple enough to achieve a good trade-off between expressiveness and complexity. For example, SCA termination is decidable and ranking functions can be extracted on terminating instances in PSPACE [4]. The simplicity of SCA sets it apart from other disjunctive abstract domains used for termination/bounds such as transition predicate abstraction [28] and powerset abstract domains [17,5].

Our method starts from the observation that progress in most software depends on the *linear change* of integer-valued functions on the program state (e.g., counter variables, size of lists, height of trees, etc.), which we call *norms*. The vast majority of non-linear bounds in real-life programs stems from two sources – nested loops and loop phases – and not from inherent non-linear behavior as in numeric algorithms. For most bounds, we have therefore the potential to exploit the nesting structure of the loops, and compose global bounds from bounds on norms. Upper bounds for norms typically consist of simple facts such as size comparisons between variables and can be computed by classical conjunctive domains. SCA is the key to convert this observation into an efficient analysis:

**(1)** Due to its built-in disjunctiveness and the transitivity of the order relations, SCA is closed under taking transitive hulls, and transitive hulls can be efficiently computed. We will use this for summarizing inner loops.

**(2)** We use SCA to compose global bounds from bounds on the norms. To extract norms from the program, we only need to consider small program parts. After the (local) extraction we have to consider only the size-change-abstracted program for bound computation.

**(3)** SCA is the natural abstract domain in connection with two program transformations – *pathwise analysis* and *contextualization* – that make imperative programs more amenable to bound analysis. Pathwise analysis is used for reasoning about complete program paths, where inner loops are overapproximated by their transitive hulls. Contextualization adds path-sensitivity to the analysis by checking which transitions can be executed subsequently. Both transformations make use of the progress in SMT solver technology to reason about the long pieces of straight-line code given by program paths.

*Summary of our Approach.* To determine how often a location $l$ of program $P$ can be visited, we proceed in two steps akin to [17]: First, we compute a disjunctive transition system $\mathcal{T}$ for $l$ from $P$. Second, we use $\mathcal{T}$ to compute a bound on the number of visits to $l$. For the first step we recursively compute transition systems for nested loops and summarize them disjunctively by transitive hulls

*Example 1.*
```
  void main (int n){
    int i = 0; int j;
l₁: while(i < n) {
      i++; j := 0;
l₂:    while((i < n) && ndet()){
         i++; j++; }
      if (j > 0)
        i--; } }
```

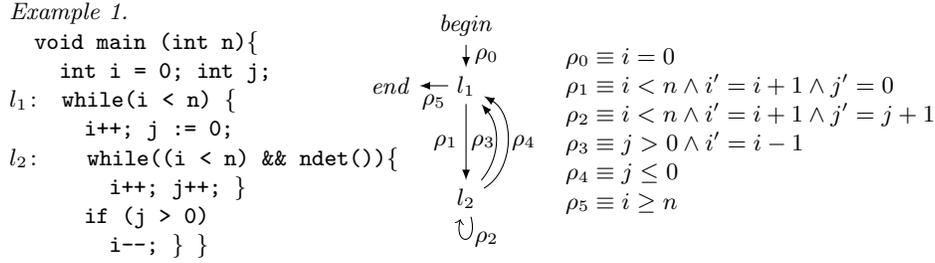

**Fig. 1.** Example 1 with its (simplified) CFG and transition relations.

computed with SCA. We enumerate all cycle-free paths from $l$ back to $l$, and derive a disjunctive transition system $\mathcal{T}$ from these paths and the summaries of the inner loops using *pathwise analysis*. For the second step we exploit the potential of SCA for automatic bound computation by first abstracting $\mathcal{T}$ using norms extracted from the program and then computing bounds solely on the abstraction. We use *contextualization* to increase the precision of the bound computation. Our method thus clearly addresses the challenges **(A)** to **(D)** discussed above. In particular, we make the following new contributions:

- We are the first to exploit SCA for bound analysis by using its ability of composing global bounds from bounds on locally extracted norms and disjunctive reasoning. Our technical contributions are the first algorithm for computing bounds with SCA (Algorithm 2) and the disjunctive summarization of inner loops with SCA (Algorithm 1).
- We are the first to describe how to apply SCA on imperative programs. Our technical contributions are two program transformations: pathwise analysis (Subsection 5.2), which exploits the looping structure of imperative programs, and contextualization (Subsection 6.1). These program transformations make imperative programs amenable to bound analysis by SCA.
- We obtain a competitive bound algorithm that captures the essential ideas of earlier termination and bound analyses in a simpler framework. Since bound analysis generalizes termination analysis, many of our methods are relevant for termination. Our experimental section shows that we can handle a large percentage of standard C programs. We give a detailed comparison with related work on termination and bound analysis in Section 8.

## 2 Examples

We use two examples to demonstrate the challenges in the automatic generation of transition systems and bound computation, and give an overview of our approach. In the examples, we denote transition relations as expressions over primed and unprimed state variables in the usual way.

*Example 1: Transition System Generation.* Let us consider the source code of Example 1 together with its (simplified) CFG and transition relations in Figure 1. Computing a bound for the header of the outer loop $l_1$ exhibits

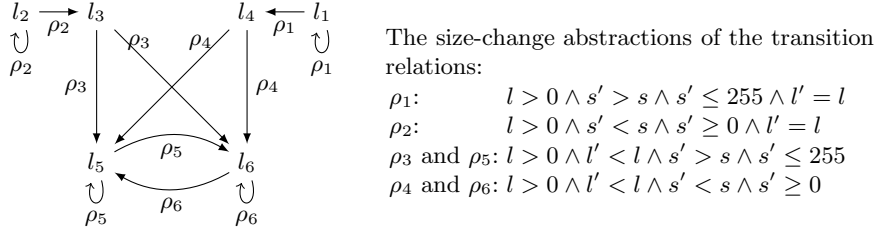

**Fig. 2.** The CFG obtained from contextualizing the transition system of Example 2 (left) and the size-change abstractions of the transition relations (right)

the following difficulties: The inner loop cannot be excluded in the analysis of the outer loop (e.g. by the standard technique called *slicing*) as it modifies the counter of the outer loop; this demonstrates the need for global reasoning in bound analysis. Further one needs to distinguish whether the inner loop has been skipped or executed at least one time as this determines whether $j = 0$ or $j > 0$. This exemplifies why we need disjunctive invariants for inner loops. Moreover, the counter $i$ may decrease, but this can only happen when $i$ has been increased by at least 2 before. This presents a difficulty to an automatic analysis since the used abstract domains need to be precise enough to capture such reasoning. In particular, a naive application of the size-change abstraction is too imprecise, since it contains only inequalities.

Our Algorithm 1 computes a transition system for the outer loop with header $l_1$ as follows: The algorithm is based on the idea of enumerating all paths from $l_1$ back to $l_1$ in order to derive a precise disjunctive transition system. However, this enumeration is not possible as there are infinitely many such paths because of the inner loop at $l_2$. Therefore Algorithm 1 recursively computes a transition system $\{i < n \wedge i' = i + 1 \wedge j' = j + 1 \wedge n' = n\}$ for the inner loop at $l_2$, and then summarizes the inner loop disjunctively by size-change abstracting its transition system to $\{n - i > 0 \wedge n' - i' < n - i \wedge j < j'\}$ (our analysis extracts the norms $n-i, j$ from the program using heuristics, cf. Section 7) and computing the reflexive transitive hull $\{n' - i' = n - i \wedge j' = j, n - i > 0 \wedge n' - i' < n - i \wedge j < j'\}$ in the abstract. (Note that we use sets of formulae to denote disjunctions of formulae.) Then Algorithm 1 enumerates all cycle-free paths from $l_1$ back to $l_1$. There are two such paths: $\pi_1 = l_1 \xrightarrow{\rho_1} l_2 \xrightarrow{\rho_3} l_1$ and $\pi_2 = l_1 \xrightarrow{\rho_1} l_2 \xrightarrow{\rho_4} l_1$. Algorithm 1 inserts the reflexive transitive hull $\mathcal{T}$ of the inner loop on the paths $\pi_1, \pi_2$ at the header of the inner loop $l_2$ and contracts the transition relations. This results in the two transition relations $\{\mathit{false}, n - i - 1 > 0 \wedge n' - i' < n - i \wedge j' > 0\}$ for $\pi_1$ (one for each disjunct of the summary of the inner loop), and $\{n - i > 0 \wedge n' - i' = n - i - 1 \wedge j' = 0, \mathit{false}\}$ for $\pi_2$. Note that for each path, *false* indicates that one transition relation was detected to be unsatisfiable, e.g. $n - i - 1 > 0 \wedge n' - i' < n - i - 1 \wedge j' > 0 \wedge j' \leq 0$ in $\pi_2$. Algorithm 1 returns the satisfiable two transitions as a transition system $\mathcal{T}$ for the outer loop.

Our Algorithm 2 size-change abstracts $\mathcal{T}$ (resulting in $\{n - i > 0 \wedge n' - i' < n - i \wedge j' > 0, n - i > 0 \wedge n' - i' < n - i \wedge j' >= 0\}$) and computes the bound

max(n, 0) from the abstraction. The difficult part in analyzing Example 1 is the transition system generation, while computing a bound from $\mathcal{T}$ is easy.

**Example 2: Bound Computation.** Bound analysis is complicated when a loop contains a finite state machine that controls its dynamics. Example 2, found during our experiments on the cBench benchmark [1], presents such a loop.

*Example 2.* // `cBench/consumer_lame/src/quantize-pvt.c`
```
int bin_search_StepSize2 (int r, int s) {
  static int c = 4; int n; int f = 0; int d = 0;
  do {
    n = nondet();
    if (c == 1 ) break;
    if (f) c /= 2;
    if (n > r) {
      if (d == 1 && !f) {f = 1; c /= 2; }
      d = 2; s += c;
      if (s > 255) break; }
    else if (n < r) {
      if (d == 2 && !f) {f = 1; c /= 2; }
      d = 1; s -= c;
      if (s < 0) break; }
    else break; }
  while (1); }
```

The loop has three different phases: in its first iteration it assigns 1 or 2 to $d$, then either increases or decreases $s$ until it sets $f$ to true; then it divides $c$ by 2 until the loop is exited. Note that disjunctive reasoning is crucial to distinguish the phases!

Our method first uses a standard invariant analysis (such as the octagon analysis) to compute the invariant $c \geq 1$, which is valid throughout the execution of the loop. Then Algorithm 1 obtains a transition system from the loop by collecting all paths from loop header back to the loop header. Omitting transitions that belong to infeasible paths we obtain six transitions:

$\rho_1 \equiv c \geq 1 \wedge \neg f \wedge d \neq 1 \wedge d' = 2 \wedge s' = s + c \wedge s' \leq 255 \wedge c' = c \wedge f' = f$
$\rho_2 \equiv c \geq 1 \wedge \neg f \wedge d \neq 2 \wedge d' = 1 \wedge s' = s - c \wedge s' \geq 0 \wedge c' = c \wedge f' = f$
$\rho_3 \equiv c \geq 1 \wedge \neg f \wedge d = 1 \wedge f' \wedge c' = c/2 \wedge d' = 2 \wedge s' = s + c' \wedge s' \leq 255$
$\rho_4 \equiv c \geq 1 \wedge \neg f \wedge d = 2 \wedge f' \wedge c' = c/2 \wedge d' = 1 \wedge s' = s - c' \wedge s' \geq 0$
$\rho_5 \equiv c \geq 1 \wedge f \wedge c' = c/2 \wedge d' = 2 \wedge s' = s + c' \wedge s' \leq 255 \wedge f' = f$
$\rho_6 \equiv c \geq 1 \wedge f \wedge c' = c/2 \wedge d' = 1 \wedge s' = s - c' \wedge s' \geq 0 \wedge f' = f$

Our bound analysis reasons about this transition system automatically by applying the program transformation called *contextualization*, which determines in which context transitions can be executed, and size-change abstracting the transitions. By our heuristics (cf. Section 7) we consider $s$ and the logarithm of $c$ (which we abbreviate by $l$) as program norms.
Figure 2 shows the CFG obtained from contextualizing the transition system of Example 2 on the left. The CFG vertices carry the information which transition is executed next. The CFG edges are labeled by the transitions of the transition system, where presence of edges indicates that, e.g., $l_4$ can be directly executed after $l_1$, and absence of an arc from $l_4$ to $l_1$ means that this transition is infeasible. The CFG shows that the transitions cannot interleave in arbitrary order; particularly useful are the strongly-connected components (SCCs) of the CFG. Our bound Algorithm 2 exploits the SCC decomposition. It computes bounds for every SCC separately using the size-change abstracted transitions (cf. Figure 2 on the right) and composes them to the overall bound $\max(255, s) + 3$, which is precise.

We point out how the above described approach *enables* automatic bound analysis by SCA. Note that the variables $d$ and $f$ do not appear in the abstracted transitions. It is sufficient for our analysis to work with the CFG obtained from contextualization because the loop behavior of Example 2, which is controlled by $d$ and $f$, has been encoded into the CFG. This has the advantage that less variables have to be considered in the actual bound analysis. Further note that the CFG decomposition gives us compositionality in bound analysis. Our analysis is able to combine the bounds of the SCCs to an (fairly complicated) overall bound using the operators max and + by following the structure of the CFG.

## 3  Program Model and Size-change Abstraction

***Sets and Relations.*** Let $A$ be a set. The concatenation of two relations $B_1, B_2 \in 2^{A \times A}$ is the relation $B_1 \circ B_2 = \{(e_1, e_3) \mid \exists e_2.(e_1, e_2) \in B_1 \wedge (e_2, e_3) \in B_2\}$. $Id = \{(e, e) \mid e \in A\}$ is the *identity relation* over $A$. Let $B \in 2^{A \times A}$ be a relation. We inductively define the *k-fold exponentiation* of $B$ by $B^k = B^{k-1} \circ B$ and $B^0 = Id$. $B^+ = \bigcup_{k \geq 1} B^k$ resp. $B^* = \bigcup_{k \geq 0} B^k$ is the *transitive-* resp. *reflexive transitive hull* of $B$. We lift the concatenation operator $\circ$ to sets of relations by defining $\mathcal{C}_1 \circ \mathcal{C}_2 = \{B_1 \circ B_2 \mid B_1 \in \mathcal{C}_1, B_2 \in \mathcal{C}_2\}$ for sets of relations $\mathcal{C}_1, \mathcal{C}_2 \subseteq 2^{A \times A}$. We set $\mathcal{C}^0 = \{Id\}$; $\mathcal{C}^k, \mathcal{C}^+$ etc. are defined analogously.

***Program Model.*** We introduce a simple program model for sequential imperative programs without procedures. Our definition models explicitly the essential features of imperative programs, namely branching and looping. In Section 5 we will explain how to exploit the graph structure of programs in our analysis algorithm. We leave the extension to concurrent and recursive programs for future work.

**Definition 1 (Transition Relations / Invariants).** *Let $\Sigma$ be a set of* states. *The set of* transition relations *$\Gamma = 2^{\Sigma \times \Sigma}$ is the set of relations over $\Sigma$. A* transition set *$\mathcal{T} \subseteq \Gamma$ is a finite set of transition relations. Let $\rho \in \Gamma$ be a transition relation. $\mathcal{T}$ is a* transition system *for $\rho$, if $\rho \subseteq \bigcup \mathcal{T}$. $\mathcal{T}$ is a* transition invariant *for $\rho$, if $\rho^* \subseteq \bigcup \mathcal{T}$.*

**Definition 2 (Program, Path, Trace, Termination).** *A* program *is a tuple $P = (L, E)$, where $L$ is a finite set of* locations, *and $E \subseteq L \times \Gamma \times L$ is a finite set of* transitions. *We write $l_1 \xrightarrow{\rho} l_2$ to denote a transition $(l_1, \rho, l_2)$.*

*A* path *of $P$ is a sequence $l_0 \xrightarrow{\rho_0} l_1 \xrightarrow{\rho_1} \cdots$ with $l_i \xrightarrow{\rho_i} l_{i+1} \in E$ for all $i$. Let $\pi = l_0 \xrightarrow{\rho_0} l_1 \xrightarrow{\rho_1} l_2 \cdots l_k \xrightarrow{\rho_k} l_{k+1}$ be a finite path. $\pi$ is* cycle-free, *if $\pi$ does* *not visit a location twice except for the end location, i.e., $l_i \neq l_j$ for all $0 \leq i < j \leq k$. The* contraction *of $\pi$ is the transition relation* $\mathtt{rel}(\pi) = \rho_0 \circ \rho_1 \circ \cdots \circ \rho_k$ *obtained from concatenating all transition relations along $\pi$. Given a location $l$, $\mathtt{paths}(P, l)$ is the set of all finite paths with start and end location $l$. A path $\pi \in \mathtt{paths}(P, l)$ is* simple, *if all locations, except for the start and end location, are different from $l$.*

*A* trace *of $P$ is a sequence $(l_0, s_0) \xrightarrow{\rho_0} (l_1, s_1) \xrightarrow{\rho_1} \cdots$ such that $l_0 \xrightarrow{\rho_0} l_1 \xrightarrow{\rho_1} \cdots$ is a path of $P$, $s_i \in \Sigma$ and $(s_i, s_{i+1}) \in \rho_i$ for all $i$. $P$ is* terminating, *if there is no infinite trace of $P$.*

Note that a cycle-free path $\pi \in \mathtt{paths}(P, l)$ is always simple. Further note that our definition of programs allows to model branching and looping precisely and naturally: imperative programs can usually be represented as CFGs whose edges are labeled with assign and assume statements.

**Definition 3 (Transition Relation of a Location).** *Let $P = (L, E)$ be a program and $l \in L$ a location. The* transition relation of $l$ *is the set $P|_l = \bigcup_{simple \ \pi \in \mathtt{paths}(P,l)} \mathtt{rel}(\pi)$.*

### 3.1 Order Constraints

Let $X$ be a set of variables. Given a variable $x$ we denote by $x'$ its *primed* version. We denote by $X'$ the set $\{x' \mid x \in X\}$ of the primed variables of $X$. We denote by $\triangleright$ any element from $\{>, \geq\}$.

**Definition 4 (Order Constraint).** *An* order constraint *over $X$ is an inequality $x \triangleright y$ with $x, y \in X$.*

**Definition 5 (Valuation).** *The set of all* valuations *of $X$ is the set $Val_X = X \to \mathbb{Z}$ of all functions from $X$ to the integers. Given a valuation $\sigma \in Val_X$ we define its* primed valuation *as the function $\sigma' \in Val_{X'}$ with $\sigma'(x') = \sigma(x)$ for all $x \in X$. Given two valuations $\sigma_1 \in Val_{X_1}, \sigma_2 \in Val_{X_2}$ with $X_1 \cap X_2 = \emptyset$ we define their* union *$\sigma_1 \cup \sigma_2 \in Val_{X_1 \cup X_2}$ by $(\sigma_1 \cup \sigma_2)(x) = \begin{cases} \sigma_1(x) & \text{for } x \in X_1, \\ \sigma_2(x) & \text{for } x \in X_2. \end{cases}$*

**Definition 6 (Semantics).** *We define a* semantic relation $\models$ *as follows: Let $\sigma \in Val_X$ be a valuation. Given an order constraint $x_1 \triangleright x_2$ over $X$, $\sigma \models x_1 \triangleright x_2$ holds, if $\sigma(x_1) \triangleright \sigma(x_2)$ holds in the structure of the integers $(\mathbb{Z}, \geq)$. Given a set $O$ of order constraints over $X$, $\sigma \models O$ holds, if $\sigma \models o$ holds for all $o \in O$.*

### 3.2 Size-change Abstraction (SCA)

We are using integer-valued functions on the program states to measure progress of a program. Such functions are called norms in the literature. Norms provide us sizes of states that we can compare. We will use norms for abstracting programs.

**Definition 7 (Norm).** *A norm $n \in \Sigma \to \mathbb{Z}$ is a function that maps the states to the integers.*

We fix a finite set of norms $N$ for the rest of this subsection, and describe in Section 7 how to extract norms from programs automatically. Given a state $s \in \Sigma$ we define a valuation $\sigma_s \in Val_N$ by setting $\sigma_s(n) = n(s)$.

We will now introduce SCA. Our terminology diverts from the seminal papers on SCA [24,4] because we focus on a logical rather than a graph-theoretic representation. The set of norms $N$ corresponds to the SCA "variables" in [24,4].

**Definition 8 (Monotonicity Constraint, Size-change Relation / Set, Concretization).** *The set of* monotonicity constraints *MCs is the set of all order constraints over $N \cup N'$. The set of* size-change relations *(SCRs) $SCRs = 2^{MCs}$ is the powerset of MCs. An SCR set $\mathcal{S} \subseteq SCRs$ is a set of SCRs. We use the* concretization function $\gamma : SCRs \to \Gamma$ *to map an SCR $T \in SCRs$ to a transition relation $\gamma(T)$ by defining $\gamma(T) = \{(s_1, s_2) \in \Sigma \times \Sigma \mid \sigma_{s_1} \cup \sigma'_{s_2} \models T\}$ as the set of all pairs of states such that the evaluation of the norms on these states satisfy all the constraints of $T$. We lift the concretization function to SCR sets by setting $\gamma(\mathcal{S}) = \{\gamma(T) \mid T \in \mathcal{S}\}$ for an SCR set $\mathcal{S}$.*

Note that the abstract domain of SCRs has only finitely many elements, namely $3^{(2|N|)^2}$. Further note that an SCR set corresponds to a formula in DNF.

**Definition 9 (Abstraction Function).** *The* abstraction function $\alpha : \Gamma \to SCRs$ *takes a transition relation $\rho \in \Gamma$ and returns the greatest SCR containing it, namely $\alpha(\rho) = \{c \in MCs \mid \rho \subseteq \gamma(c)\}$. We lift the abstraction function to transition sets by setting $\alpha(\mathcal{T}) = \{\alpha(\rho) \mid \rho \in \mathcal{T}\}$ for a transition set $\mathcal{T}$.*

*Implementation of the abstraction.* $\alpha$ can be implemented by an SMT solver under the assumption that the norms are provided as expressions and that the transition relation is given as a formula such that the order constraints between these expressions and the formula fall into a logic that the SMT solver can decide.

Using abstraction and concretization we can define concatenation of SCRs:

**Definition 10 (Concatenation of SCRs).** *Given two SCRs $T_1, T_2 \in SCRs$, we define $T_1 \circ T_2$ to be the SCR $\alpha(\gamma(T_1) \circ \gamma(T_2))$. We lift the concatenation operator $\circ$ to SCR sets by defining $\mathcal{S}_1 \circ \mathcal{S}_2 = \{T_1 \circ T_2 \mid T_1 \in \mathcal{S}_1, T_2 \in \mathcal{S}_2\}$ for SCR sets $\mathcal{S}_1, \mathcal{S}_2 \in 2^{SCRs}$. $\mathcal{S}^0 = \{Id\}, \mathcal{S}^k, \mathcal{S}^+, \mathcal{S}^*$ etc. are defined in the natural way.*

Concatenation of SCRs is conservative by definition, i.e., $\gamma(T_1 \circ T_2) \supseteq \gamma(T_1) \circ \gamma(T_2)$ and associative because of the transitivity of order relations. Concatenation of SCRs can be effectively computed by a modified all-pairs-shortest-path algorithm (taking order relations as weights). Because the number of SCRs is finite, the transitive hull is computable.

The following theorem can be directly shown from the definitions. We will use it to summarize the transitive hull of loops disjunctively, cf. Section 5.

**Theorem 1 (Soundness).** *Let $\rho$ be a transition relation and $\mathcal{T}$ a transition system for $\rho$. Then $\gamma(\alpha(\mathcal{T})^*)$ is a transition invariant for $\rho$.*

## 4 Main Steps of our Analysis

Let $P = (L, E)$ be a program and $l \in L$ be a location for which we want to compute a bound. Our analysis consists of four main steps:

| | |
|---|---|
| 1. Extract a set of norms $N$ using heuristics | (Section 7) |
| 2. Compute global invariants by standard abstract domains | |
| 3. Compute $\mathcal{T} = \texttt{TransSys}(P, l)$ | (Section 5) |
| 4. Compute $b = \texttt{Bound}(\texttt{Contextualize}(\mathcal{T}))$ | (Section 6) |

In Step 1 we extract a set of norms $N$ using the heuristics described in Section 7. The abstraction function $\alpha$ that we use in Steps 3 and 4 is parameterized by the set of norms $N$. In Step 2 we compute global invariants by standard abstract domains such as interval, octagon or polyhedra. As this step is standard, we do not discuss it in this paper. In Step 3 we compute a transition system $\mathcal{T} = \texttt{TransSys}(P, l)$ for $P|_l$ by Algorithm 1. In Step 4 we compute a bound $b = \texttt{Bound}(\texttt{Contextualize}(\mathcal{T}))$ for the number of visits to $l$, where we first use the program transformation contextualization of Definition 11 to transform $\mathcal{T}$ into a program from which we then compute a bound $b$ by Algorithm 2.

---

**Procedure**: $\texttt{TransSys}(P, l)$
**Input**: a program $P = (L, E)$, a location $l \in L$
**Output**: a transition system for $P|_l$
**Global**: array summary for storing transition invariants

**foreach** $(loop, header) \in NestedLoops(P, l)$ **do**
$\quad \mathcal{T} := \texttt{TransSys}(loop, header);$
$\quad hull := \gamma(\alpha(\mathcal{T})^*);$
$\quad \textsf{summary}[header] := hull;$

**foreach** cycle-free path $\pi = l \xrightarrow{\rho_0} l_1 \xrightarrow{\rho_1} l_2 \cdots l_k \xrightarrow{\rho_k} l \in \texttt{paths}(P, l)$ **do**
$\quad \mathcal{T}_\pi := \{\rho_0\} \circ \texttt{ITE}(\texttt{IsHeader}(l_1), \textsf{summary}[l_1], \{Id\}) \circ \{\rho_1\} \circ$
$\quad \quad \quad \texttt{ITE}(\texttt{IsHeader}(l_2), \textsf{summary}[l_2], \{Id\}) \circ \{\rho_2\} \circ \cdots \circ$
$\quad \quad \quad \texttt{ITE}(\texttt{IsHeader}(l_k), \textsf{summary}[l_k], \{Id\}) \circ \{\rho_k\};$

**return** $\bigcup_{\text{cycle-free path } \pi \in \texttt{paths}(P, l)} \mathcal{T}_\pi;$

**Algorithm 1**: $\texttt{TransSys}(P, l)$ computes a transition system for $P|_l$

## 5 Computing Transition Systems

In this section we describe our algorithm for computing transition systems. We first present the actual algorithm, and then discuss specific characteristics. The function $\texttt{TransSys}$ in Algorithm 1 takes as input a program $P = (L, E)$ and a location $l \in L$ and computes a transition system for $P|_l$, cf. Theorem 2 below. The key ideas of Algorithm 1 are (1) to summarize inner loops disjunctively by transition invariants computed with SCA, and (2) to enumerate all cycle-free paths for pathwise analysis. Note that for loop summarization the algorithm is recursively invoked. We give an example for the application of Algorithm 1 to Example 1 in Section B.1 of the appendix.

***Loop Summarization.*** In the first `foreach`-loop, Algorithm 1 iterates over all nested loops of $P$ w.r.t. $l$. A loop *loop* of $P$ is a nested loop w.r.t. $l$, if it is strongly connected to $l$ but does not contain $l$, and if there is no loop with the same properties that strictly contains *loop*. Let *loop* be a nested loop of $P$ w.r.t. $l$ and let *header* be its header. (We assume that the program is reducible, see discussion below.) `TransSys` calls itself recursively to compute a transition system $\mathcal{T}$ for $loop|_{header}$.

In statement $hull := \gamma(\alpha(\mathcal{T})^*)$, $\alpha(\mathcal{T})$ size-change abstracts $\mathcal{T}$ to an SCR set, $\alpha(\mathcal{T})^*$ computes the transitive hull of this SCR set, and $\gamma(\alpha(\mathcal{T})^*)$ concretizes the abstract transitive hull to a transition set, which is then assigned to *hull*. Algorithm 1 stores *hull* in the array `summary`, which is a transition invariant for $loop|_{header}$ by the soundness of SCA as stated in Theorem 1.

After the first `foreach`-loop, Algorithm 1 has summarized all inner loops, not only the nested loops, because the recursive calls reaches all nesting levels. For each inner loop *loop* with header *header* a transition invariant for $loop|_{header}$ has been stored at `summary`[*header*]. Summaries of inner loops are visible to all outer loops, because the array `summary` is a global variable.

***Pathwise Analysis.*** In the second `foreach`-loop, Algorithm 1 iterates over all cycle-free paths of $P$ with start and end location $l$. Let $\pi = l \xrightarrow{\rho_0} l_1 \xrightarrow{\rho_1} \cdots l_k \xrightarrow{\rho_k} l$ be such a cycle-free path. The expression $\texttt{ITE}(\texttt{IsHeader}(l_i), \textsf{summary}[l_i], \{Id\})$ evaluates to summary$[l_i]$ for each location $l_i$, if $l_i$ is the header of an inner loop $loop_i$, and evaluates to the transition set $\{Id\}$, which contains only the identity relation over the program states, else. Algorithm 1 computes the set $\mathcal{T}_\pi = \{\rho_0\} \circ \texttt{ITE}(\texttt{IsHeader}(l_1), \textsf{summary}[l_1], \{Id\}) \circ \{\rho_1\} \circ \texttt{ITE}(\texttt{IsHeader}(l_2), \textsf{summary}[l_2], \{Id\}) \circ \{\rho_2\} \circ \cdots \circ \texttt{ITE}(\texttt{IsHeader}(l_k), \textsf{summary}[l_k], \{Id\}) \circ \{\rho_k\}$, which is an overapproximation of the contraction of $\pi$, where the summaries of the inner loops $loop_i$ are inserted at their headers $l_i$. The transition set $\mathcal{T}_\pi$ overapproximates all paths starting and ending in $l$ that iterate arbitrarily often through inner loops along $\pi$, because for every loop $loop_i$ the transition set summary$[l_i]$ overapproximates all paths starting and ending in $l_i$ that iterate arbitrarily often through $loop_i$ (as summary$[l_i]$ is a transition invariant for $loop_i|_{l_i}$). Algorithm 1 returns the union $\bigcup_{\text{cycle-free path } \pi \in \texttt{paths}(P,l)} \mathcal{T}_\pi$ of all those transition sets $\mathcal{T}_\pi$.

**Theorem 2.** *Algorithm 1 computes a transition system* $\texttt{TransSys}(P, l)$ *for* $P|_l$.

*Proof (Sketch).* Let $\pi' \in \texttt{paths}(P, l)$ be a simple path. We obtain a cycle-free path $\pi \in \texttt{paths}(P, l)$ from $\pi'$ by deleting all iterations through inner loops of $(P, l)$ from $\pi'$. The transition set $\mathcal{T}_\pi$ overapproximates all paths starting and ending in $l$ that iterate arbitrarily often through inner loops of $(P, l)$ along $\pi$. As $\pi'$ iterates through inner loops of $(P, l)$ along $\pi$ we have $\texttt{rel}(\pi) \subseteq \bigcup \mathcal{T}_\pi$.

*Implementation.* We use conjunctions of formulae to represent individual transitions. This allows us to implement the concatenation of transition relations by conjoining their formulae and introducing existential quantifiers for the intermediate variables. We detect empty transition relations by asking an SMT solver whether their corresponding formulae are satisfiable. We use these emptiness checks at several points during the analysis to reduce the number of transition relations.

Algorithm 1 may exponentially blow up in size because of the enumeration of all cycle-free paths and the computation of transitive hulls of inner loops. We observed in our experiments that by first extracting norms from the program under scrutiny and then slicing the program w.r.t. these norms before continuing with the analysis normally results into programs small enough for making our analysis feasible.

*Irreducible programs.* Algorithm 1 refers to loop headers, and thus implicitly assumes that loops are reducible. (Recall that in a reducible program each SCC has a unique entry point called the header.) We have formulated Algorithm 1 in this way to make clear how it exploits the looping structure of imperative programs. However, Algorithm 1 can be easily extended to irreducible loops by a case distinction on the (potentially multiple) entry points of SCCs.

### 5.1 Disjunctiveness in Algorithm 1

Disjunctiveness is crucial for bound analysis. We have given two examples for this fact in Section 2 and refer the reader for further examples to [17,5,27]. We emphasize that our analysis can handle *all examples* of these publications, and give a detailed comparison with them in Section 8. Our analysis is disjunctive in two ways:

(1) *We summarize inner loops disjunctively.* Given a transition system $\mathcal{T}$ for some inner loop *loop*, we want to summarize *loop* by a transition invariant. The most precise transition invariant $\mathcal{T}^* = \{Id\} \cup \mathcal{T} \cup \mathcal{T}^2 \cup \mathcal{T}^3 \cup \cdots$ introduces infinitely many disjunctions and is not computable in general. In contrast to this the abstract transitive hull $\alpha(\mathcal{T})^* = \alpha(\{Id\}) \cup \alpha(\mathcal{T}) \cup \alpha(\mathcal{T})^2 \cup \alpha(\mathcal{T})^3 \cup \cdots$ has only finitely many disjunctions and is effectively computable. This allows us to overapproximate the infinite disjunction $\mathcal{T}^*$ by the finite disjunction $\gamma(\alpha(\mathcal{T})^*)$.

We underline that the need for disjunctive summaries of inner loops in the bound analysis is a major motivation for SCA, as it allows us to compute disjunctive transitive hulls naturally, cf. definition and discussion in Section 3.2.

(2) *We summarize local transition relations disjunctively.* Given a program $P = (L, E)$ and location $l \in L$, we want to compute a transition system for $P|_l$. For a cycle-free path $\pi \in \texttt{paths}(P, l)$ the transition set $\mathcal{T}_\pi$ computed in Algorithm 1 overapproximates all simple paths in $\texttt{paths}(P, l)$ that iterate through inner loops along $\pi$. As all $\mathcal{T}_\pi$ are sets, the set union $\bigcup_{\text{cycle-free path } \pi \in \texttt{paths}(P,l)} \mathcal{T}_\pi$ is a disjunctive summarization of all $\mathcal{T}_\pi$ that keeps the information from different paths separated. This is important for our analysis which relies on the observation that monotonic changes of norms can be observed along single paths from loop header back to the header.

### 5.2 Pathwise Analysis in Algorithm 1

It is well-known that analyzing large program parts jointly improves the precision of static analyses, e.g. [7]. Owing to the progress in SMT solvers this idea has recently seen renewed interested by static analyses such as abstract interpretation [26] and software model checking [6], which use SMT solvers for abstracting large blocks of straight-line code jointly to increase the precision of the analysis.

We call the analyses of [26,6] and classical SCA [24,4] *blockwise*, because they do joint abstraction only for loop-free program parts. In contrast, our *pathwise analysis* abstracts complete paths at once: Algorithm 1 enumerates all cycle-free paths from loop header to loop header and inserts summaries for inner loops on these paths. These paths are then abstracted jointly in a subsequent loop summarization or bound computation. In this way our pathwise analysis is strictly more precise than blockwise analysis. We illustrate this difference in precision on Example 1 in Section A of the appendix.

Parsers are a natural class of programs which illustrate the need for pathwise analysis. In our experiments we observed that many parsers increase an index while scanning the input stream and use lookahead to detect which token comes next. As in Example 1, lookaheads may temporarily decrease the index. Pathwise abstraction is crucial to reason about the program progress with SCA.

## 6 Bound Computation

Our bound computation consists of two steps. Step 1 is the program transformation contextualization which transforms a transition system into a program. Step 2 is the bound algorithm which computes bounds from programs.

### 6.1 Contextualization

Contextualization is a program transformation by Manolios and Vroon [25], who report on an impressive precision of their SCA-based termination analysis of functional programs. Note that we do not use their terminology (e.g. "calling context graphs") in this paper. Our contribution lies in adopting contextualization to imperative programs and in recognizing its relevance for bound analysis.

**Definition 11 (Contextualization).** *Let $\mathcal{T}$ be a transition set. The contextualization of $\mathcal{T}$ is the program $P = (\mathcal{T}, E)$, where $E = \{\rho \xrightarrow{\rho} \rho' \mid \rho, \rho' \in \mathcal{T} \text{ and } \rho \circ \rho' \neq \emptyset\}$.*

The contextualization of a transition system is a program in which every location determines which transition is executed next; the program has an edge between two locations only if the transitions of the locations can be executed one after another.

Contextualization restricts the order in which the transitions of the transition system can be executed. Thus, contextualization encodes information that could otherwise be deduced from the pre- and postconditions of transitions directly into the CFG. Since pathwise analysis contracts whole loop paths into single transitions, contextualization is particularly important after pathwise analysis: our subsequent bound algorithm does not need to compute the pre- and postcondition of the contracted loop paths but only needs to exploit the structure of the CFGs for determining in which order the loop paths can be executed.

*Example 3.*
```
void main (int x, int b){
  while (0 < x < 255){
    if (b) x = x + 1;
    else x = x - 1; } }
```

$\rho_1 \equiv 0 < x < 255 \land b \land x' = x + 1 \land b'$
$\rho_2 \equiv 0 < x < 255 \land \neg b \land x' = x - 1 \land \neg b'$

$l_1 \quad l_2$
$\circlearrowleft \quad \circlearrowleft$
$\rho_1 \quad \rho_2$

**Fig. 3.** Example 3 with its transition relations and CFG obtained from contextualization.

We illustrate contextualization on Example 3. The program has two paths, and gives rise to the transition system $\mathcal{T} = \{\rho_1, \rho_2\}$. Keeping track of the boolean variable $b$ is important for bound analysis: Without reference to $b$ not even the termination of main can be proven. In Figure 3 (right) we show the contextualization of $\mathcal{T}$. Note that contextualization has encoded information about the variable $b$ into the CFG in such a way that we do not need to keep track of the variable $b$ anymore. Thus, contextualization releases us from taking the precondition $b$ resp. $\neg b$ and the postcondition $b'$ resp. $\neg b'$ into account for bound analysis.

At the beginning we gave an application of contextualization on the sophisticated loop in Example 2, where contextualization uncovers the control structure of the finite state machine encoded into the loop. An application of contextualization to the flagship example of a recent publication [14] can be found in Section B.3 of the appendix.

Note that in our definition of contextualization we only consider the consistency of two consecutive transitions. It would also have been possible to consider three or more consecutive transitions. This would result in increased precision. However, we found two transitions to be sufficient in practice.

*Implementation.* We implement contextualization by encoding the concatenation $\rho_1 \circ \rho_2$ of two transitions $\rho_1, \rho_2$ into a logical formula and asking an SMT solver whether this formula is satisfiable. Note that such a check is very simple to implement in comparison to the explicit computation of pre- and postconditions.

---

**Procedure**: Bound($P$)
**Input**: a program $P = (L, E)$
**Output**: a bound $b$ on the length of the traces of $P$
$SCCs := \texttt{computeSCCs}(P); b := 0;$
**while** $SCCs \neq \emptyset$ **do**
    $SCCsOnLevel := \emptyset;$
    **forall the** $SCC \in SCCs$ *s.t. no* $SCC' \in SCCs$ *can reach* $SCC$ **do**
        $r := \texttt{BndSCC}(SCC);$
        Let $r \leq b_{SCC}$ be a global invariant;
        $SCCsOnLevel := SCCsOnLevel \cup \{SCC\};$
    $b := b + \max_{SCC \in SCCsOnLevel} b_{SCC};$
    $SCCs := SCCs \setminus SCCsOnLevel;$
**return** $b;$

**Algorithm 2:** Bound composes the bounds of the SCCs to an overall bound

## 6.2 Bound Algorithm

Our bound algorithm reduces the task of bound computation to the computation of local bounds and the composition of these local bounds to an overall bound. To this end, we exploit the structure of the CFGs obtained from contextualization: We partition the CFG of programs into its strongly connected components (SCCs) (SCCs are maximal strongly connected subgraphs). For each SCC, we compute a bound by Algorithm 3, and then compose these bounds to an overall bound by Algorithm 2.

Algorithm 2 arranges the SCCs of the CFG into levels: The first level consists of the SCCs that do not have incoming edges, the second level consists of the SCCs that can be reached from the first level, etc. For each level, Algorithm 2 calls Algorithm 3 to compute bounds for the SCCs of this level. Let $SCC$ be an SCC of some level and let $r := \texttt{BndSCC}(SCC)$ be the bound returned by Algorithm 3 on $SCC$. $r$ is a (local) bound of $SCC$ that may contain variables of $P$ that are changed during the execution of $P$. Algorithm 2 uses global invariants (e.g. interval, octagon or polyhedra) in order to obtain a bound $b_{SCC}$ on $r$ in terms of the initial values of $P$. The SCCs of one level are collected in the set $SCCsOnLevel$. For each level, Algorithm 2 composes the bounds $b_{SCC}$ of all SCCs $SCC \in SCCsOnLevel$ to a maximum expression. Algorithm 2 sums up the bounds of all levels for obtaining an overall bound.

---

**Procedure**: $\texttt{BndSCC}(P)$
**Input**: strongly-connected program $P = (L, E)$
**Output**: a bound $b$ on the length of the traces of $P$
**if** $E = \emptyset$ **then return** 1;
$NonIncr := \emptyset$; $DecrBnded := \emptyset$; $BndedEdgs := \emptyset$;
**foreach** $n \in N$ **do**
  **if** $\forall\ l_1 \xrightarrow{\rho} l_2 \in E\ n \geq n' \in \alpha(\rho)$ **then**
    $NonIncr := NonIncr \cup \{n\}$;

**foreach** $l_1 \xrightarrow{\rho} l_2 \in E,\ n \in NonIncr$ **do**
  **if** $n \geq 0, n > n' \in \alpha(\rho)$ **then**
    $DecrBnded := DecrBnded \cup \{\max(n, 0)\}$;
    $BndedEdgs := BndedEdgs \cup \{l_1 \xrightarrow{\rho} l_2\}$;

**if** $BndedEdgs = \emptyset$ **then fail with** "there is no bound for $P$";
$b = \texttt{Bound}((L, E \setminus BndedEdgs))$;
**return** $((\sum DecrBnded) + 1) \cdot b$;

**Algorithm 3:** $\texttt{BndSCC}$ computes a bound for a single SCC

---

Algorithm 3 computes the bound of a strongly-connected program $P$. First Alg. 3 checks if $P = (L, E)$ is trivial, i.e., $E = \emptyset$, and returns 1, if this is the case. Next Alg. 3 collects all norms in the set *NonIncr* that either decrease or stay equal on all transitions. Subsequently Alg. 3 checks for every norm $n \in NonIncr$ and transition $l_1 \xrightarrow{\rho} l_2 \in E$, if $n$ is bounded from below by zero and decreases on $\rho$. If this is the case, Alg. 3 adds $\max(n, 0)$ to the set *DecrBnded* and $l_1 \xrightarrow{\rho} l_2$ to *BndedEdgs*. Note that the transitions included in the set *BndedEdgs* can only be executed as long as their associated norms are greater than zero. Every transition in *BndedEdgs* decreases an expression in *DecrBnded* when it is taken. As the expressions in *DecrBnded* are never increased, the sum of all expressions in *DecrBnded* is a bound on how often the transitions in *BndedEdgs* can be taken. If *DecrBnded* is empty, Alg. 2 fails, because the absence of infinite cycles could not be proven. Otherwise we recursively call Alg. 2 on $(L, E \setminus \textit{BndedEdgs})$ for a bound $b$ on this subgraph. The subgraph can at most be entered as often as the transitions in *BndedEdgs* can be taken plus one (when it is entered first). Thus, $((\sum \textit{DecrBnded}) + 1) \cdot b$ is an upper bound for $P$.

*Role of SCA in our Bound Analysis.* Our bound analysis uses the size-change abstractions of transitions to determine how a norm $n$ changes according to $n \geq n'$, $n > n'$, $n \geq 0$ in Alg. 3. We plan to incorporate inequalities between different norms (like $n \geq m'$) in future work to make our analysis more precise.

*Termination analysis.* If in Algorithm 2 the global invariant analysis cannot infer an upper bound on some local bound, the algorithm fails to compute a bound, but we can still compute a lexicographic ranking function, which is sufficient to prove termination. The respective adjustment of our algorithm is straightforward.

We give an example for the application of Algorithm 2 to Example 2 and to the flagship example of [14] in Section B.2 and B.3 of the appendix.

## 7 Heuristics for Extracting Norms

In this section we describe our heuristic for extracting norms from programs. Let $P = (L, E)$ be a program and $l \in L$ be a location. We compute all cycle-free paths from $l$ back to $l$. For all arithmetic conditions $x \geq y$ appearing in some of these paths we take $x - y$ as a norm if $x - y$ decreases on this path; this can be checked by an SMT solver. Note that in such a case $x - y$ is a local ranking function for this program path. Similar patterns and checks can be implemented for iterations over bitvectors and data structures. For a more detailed discussion on how to extract the local ranking functions of a program path we refer the reader to [17]. We also compute norms for inner loops on which already extracted norms are control dependent and add them to the set of norms until a fixed point is reached (similar to program slicing). We also include the sum and the difference of two norms, if an inner loop affects two norms at the same time. Further, we include the rounded logarithm of a norm, if the norm is multiplied by a constant on some program path. In general any integer-valued expression can be used as a program norm, if considered useful by some heuristic. Clearly the analysis gets more precise the more norms are taken into account, but also more costly.

## 8 Detailed Comparison with Related Work

In this section we give a detailed comparison with earlier termination / bound analyses. We show that our bound analysis captures the essential ideas of these approaches in a simpler framework.

### 8.1 Comparison of transition predicate abstraction (TPA) and SCA by Heizmann et al.

In [18] Heizmann et al. state that SCA is an instance of the more general technique of TPA [28]. In particular they formally show that when a tail-recursive functional program $F$ is translated into an imperative program $P$, then an SCA-based termination analysis on $F$ can be mimicked by a TPA-based termination analysis on $P$ whose predicates are order relations.

However, [18] does not

- deal with general imperative programs, but with programs obtained as translations from functional programs. Since functional programs can be size-change abstracted more easily (as explained in our comparison with SCA in Section 8.7), this problem setting is much simpler.
- show how to obtain transition predicates for the independent analysis of imperative programs by TPA. It only shows that (as a result of the translation) TPA is more general than SCA.
- deal with a concrete programming language, and does not deal with practical issues, or concrete analysis tools.
- make use of the recent progress of the SCA [4], where SCA is extended from natural numbers to integers, and deals only with natural numbers.

Our paper fills in all these left open gaps. Moreover, our paper clearly goes beyond the issues discussed in [18] by unifying much of the previous work on termination and bound analysis, e.g., see our comparison to TERMINATOR in Section 8.2, SPEED in Section 8.6 etc.

While we find it quite intuitive that SCA as well as our more general approach are instances of TPA, we are concerned with a different issue in this paper. We argue that precisely because of its limited expressiveness SCA is suitable for bound analysis: abstracted programs are simple enough that we can compute bounds for them. We have shown that imperative programs are amenable to bound analysis by SCA using appropriate program transformations, whereas [18] is not concerned with practical issues.

### 8.2 Termination Analysis by TERMINATOR

The TERMINATOR tool [8] is an automatic termination analyzer of imperative programs, which uses TPA [28] for constructing a Ramsey based termination argument [27].

Our approach and TERMINATOR share the idea of extracting progress measures locally (norms resp. local ranking functions) and composing them for a global analysis (bound resp. termination proof). Because of its non-constructive nature, the Ramsey based termination argument underlying TERMINATOR cannot be used for extracting a global ranking function out of the termination proof. In contrast, we use SCA for the first time to compose global bounds from bounds on norms. Earlier work on SCA [4] already has shown how to compute global ranking functions from norms.

In order to apply the Ramsey based termination argument, TERMINATOR needs to analyze the transitive hull of programs. This analysis is the most expensive step in the analysis of TERMINATOR and has to be repeated many times. In contrast, we abstract programs first and then analyze only the transitive hull of the abstract program. This has huge benefits for the speed of the analysis (further discussed in Section 8.3).

TPA can lose precision in every step of the analysis. In contrast, our pathwise analysis follows the structure of programs and loses precision only at well-defined places. Our analysis handles paths precisely by conjoining the formulae of the statements along the path (using all theories that can be handled by SMT solvers) and loses precision only when summarizing loops. However, we handle loops precisely w.r.t. their monotonic behavior of norms: SCA is closed under taking transitive hulls because of the built-in disjunctiveness of SCA and the transitivity of the order relations, transitive hulls can be computed effectively.

TERMINATOR is built on top of a full-fledged software model checker, which implements a complicated CEGAR loop in order to extract predicates and local ranking functions from programs. In contrast, our simple and lightweight static analysis relies only on an SMT solver, our set of transition predicates is fixed in advance (the monotonicity predicates of SCA) and our set of norms is extracted from the program at the beginning of the analysis. It is an interesting direction of future work to investigate how to combine these approaches, e.g., by using coarse abstractions for filtering the "easy cases" and refining the precision for handling the "hard cases".

### 8.3 Termination Analysis by LOOPFROG

[22,30] observe that TPA-based approaches such as TERMINATOR [8] spend almost all time in analyzing the transitive hulls of programs, i.e., the expensive step is proving $P|_l^+ \subseteq \bigcup \mathcal{T}$ for transition sets $\mathcal{T}$. Therefore [22,30] take a different approach and give algorithms that search for a transitive transition system $\mathcal{T}$ for $P|_l$. A transition set $\mathcal{T}$ is *transitive*, if $\bigcup \mathcal{T}^2 \subseteq \bigcup \mathcal{T}$. A transitive transition system $\mathcal{T}$ for $P|_l$ already implies $P|_l^+ \subseteq \bigcup \mathcal{T}^+ \subseteq \bigcup \mathcal{T}$ by the transitivity of $\mathcal{T}$. This has the advantage that the expensive direct proof of $P|_l^+ \subseteq \bigcup \mathcal{T}$ is avoided.

The first version of LOOPFROG [22] implements an algorithm that constructs such transitive transition systems iteratively. In every step LOOPFROG adds transition relations to a candidate transition set $\mathcal{T}$. We argue that the effectiveness of such an iterative algorithm is limited, and that what the authors of [22] really want is SCA!

Note that a transitive transition system $\mathcal{T}$ for $P|_l$ that is precise enough to prove the termination of $P$ is an overapproximation of $P$ that still terminates.

Let us consider an example transition set $\mathcal{T} = \{\rho_1, \rho_2\}$ with $\rho_1 = x > 0 \land x > x'$ and $\rho_2 = y > 0 \land y > y'$. $\mathcal{T}$ does not terminate because $\rho_1$ can increase the value of $y$ arbitrarily and $\rho_2$ can increase the value of $x$ arbitrarily. Let us assume that the analyzed program $P$ nevertheless terminates because the variable $x$ can only be decreased when $y$ stays constant. Let us further assume that LOOPFROG has added $\rho_1$ to $\mathcal{T}$ in the first step and $\rho_2$ in the second step of its iteration. LOOPFROG could have added the information about $x$ in the first step by setting $\rho_1 = x > 0 \land x > x' \land y = y'$, but not in the second step. Note that once the candidate transition set $\mathcal{T}$ does not terminate, it cannot be repaired by adding transition relations. Note further that in later steps the loss of information of earlier steps cannot be repaired as we have seen on the above example. Thus, we conclude that a candidate transition set $\mathcal{T}$ has to be constructed in one single step. This is exactly what we do in our analysis. We first compute transition system for programs, and then size-change abstract the transition relations for our bound analysis. Alternatively, these abstracted transition relations could be analyzed for termination by a transitive hull computation using the termination criterion of SCA [4]. This transitive hull then provides exactly the transitive transition system $\mathcal{T}$ for $P|_l$. From this we conclude that what the authors of [22] really want is SCA!

The second version of LOOPFROG [30] uses relational loop summarization (see also the next subsection). For this summarization [30] uses template invariants. Only one of these templates contains disjunction (two disjuncts). [30] states that these templates are inspired by the more general size-change abstract domain. We show in this paper how to employ the full SCA domain by using pathwise analysis for exploiting the looping structure of imperative programs. This allows us to use the full disjunctive power of SCA. [30] is only concerned with termination analysis, whereas we show how to use SCA for the more difficult problem of bound analysis.

### 8.4 Loop Summarization

Loop summarization as in Algorithm 1 is being recognized as important tool in program analysis, for example [21] summarizes loops by overapproximations of the reachable states for automatic proofs of safety properties. Relational summarizations of loops have for the first time been used in the bound analysis of [17]. The termination analysis [30], which is an extension of [21], also uses relational summaries of loops.

Loop summarization is closely related to procedure summarization, e.g. [16].

### 8.5 Disjunctive Abstract Domains

The papers [17,5] on bound and termination analysis use abstract interpretation for computing disjunctive transition invariants. Both papers suggest lifting a conjunctive abstract domain $D$ to the powerset domain $2^D$, and refer to the standard octagon / polyhedra domains as instantiations of $D$ in their implementation sections.

However, lifting the octagon / polyhedra domain to the powerset domain is difficult because it requires an adequate lifting of the join operator: Set union is the precise join operator of both powerset lattices, but every set union may increase the number of the base elements of the powerset elements. Note that both powerset lattices have infinite width. Thus using set union as join operator does not guarantee the termination of the analysis. A standard way of ensuring the finiteness of the analysis is by limiting the number of base elements of a powerset element, such that every time the number of base elements of a powerset element is over the limit some base elements are merged.

[17] states an algorithm for lifting the join operator to the powerset domain based on an assumption resembling convex theories that uses the same syntactic merging function in every step of the fixed point computation. However, [17] proposes to choose the merging function heuristically or to try all merging functions for a small number of disjuncts and take the best result. [5] remains vague: "For our present empirical evaluation we use an extraction method after the fixed-point analysis has been performed in order to find disjunctive invariance/variance assertions.".

In contrast, SCA is a finite powerset domain that naturally handles disjunction: SCA has finite width, therefore we never have to merge abstract elements and can handle disjunction precisely. This releases us from relying on complicated merging algorithms as [17,5]. SCA has finite height, therefore we do not need widening to compute fixed points (e.g. transitive hulls). This releases us from lifting the widening operator of conjunctive domains (e.g. octagon, polyhedra) to powerset domains.

### 8.6 Bound Analysis by the SPEED project

In earlier work [17] we have stated proof rules for computing global bounds from local bounds of transitions. Ad hoc proof rules as in [17] often give rise to efficient analyses, but do not establish a general theory. This is unsatisfying because such a theory is crucial for investigating the completeness of the analysis and the applicability to related problems or other programming paradigms. In this paper we have identified SCA as a suitable abstraction for bound analysis. SCA provides the theory that we have been looking for, because all the proof rules of [17] are instantiations of our more general bound algorithm.

In [17] we stated a so-called *enabledness* check that detects non-interference between transitions, which can be used in the composition of the global bound. Unfortunately, this check is flawed when more than two transitions are considered. Our program transformation contextualization can soundly detect non-interference when an arbitrary number of transitions is considered.

[14] proposes to use program transformation before performing bound analysis. The program transformation stated in [14] is parameterized by an abstract domain, which is used simultaneously with the actual transformation algorithm to detect the infeasibility of certain paths. However, [14] is vague about what abstract domains should be used, and the actual transformation algorithm is quite involved. In contrast, we propose two simple program transformations that are

easy to implement. Our program transformations only rely on SMT solver calls and do not require additional abstract domains.

### 8.7 Size-change Abstraction

Despite its success in functional/declarative languages, e.g. [25], [20], SCA [24,4] has not yet been applied to imperative programs. We describe the two main obstacles in the application of SCA to imperative programs and how we solve them: (1) In functional / declarative languages, algorithms typically operate on algebraic data structures where constructs and destructs happen in single steps. Due to this succinctness, SCA achieves sufficient precision on small program blocks. In imperative programs loops can have many intermediate stages and oftentimes only the program state at the loop header can be considered as "clean". Therefore the abstraction of small program blocks to size-change relations loses too much precision. (This issue is well-known in the field of invariant computation.) We solve this issue by our pathwise analysis, which has the effect that large pieces of code that lie between the "clean" program locations are abstracted jointly. (2) The intended use of the SCA variables is as *local progress measures of the program.* In functional/declarative languages there is a natural set of such local progress measures such as the size of a data type, the height of a tree, the length of a list, or any arithmetic expression built up from those. In imperative programs, it is less clear what the shape of this local progress measures is and how they can be automatically extracted from programs. We give a solution to this problem by extracting norms from the conditions of complete loop paths (as described in our heuristics).

### 8.8 Other Approaches

A series of works describes a type-based potential-method of amortized analysis for the estimation of resource usage in first-order functional programs, which reduces the problem to linear constraint solving. Recent enhancement includes the extension to multivariate polynomial bounds [2] and higher-order programs [19].

The embedded and real-time systems community has taken considerable effort on worst case execution time (WCET) estimation [31]. WCET presents an orthogonal line of research, which for establishing loop bounds either requires user annotations or employs simple techniques based on pattern matching and numerical analysis. We report on a WCET benchmark in our experiments.

## 9 Experiments

Our tool LOOPUS applies the methods of this paper to obtain upper bounds on loop iterations. The tool employs the LLVM compiler framework and performs its analysis on the LLVM intermediate representation [23]. We are using ideal integers in our analysis instead of exact machine representation (bitvectors). Our analysis operates on the SSA variables generated by the mem2reg pass and handles memory references using optimistic assumptions. For logical reasoning we use the *yices* SMT solver [9]. Our experiments were performed on an Intel Xeon CPU (4 cores with 2.33 GHz) with 16 GB Ram.

The Mälardalen benchmark is used in the area of *worst case execution time analysis* for the comparison and evaluation of methods and tools. It contains 7497 lines of code and 262 loops. In less than 35 seconds total time, we computed a bound for 93% of the loops. On the loops with more than one path (in the following called *non-trivial loops*) we had a success ratio of 72% (42 of 58 loops). The failure cases had the following reasons: (1) unimplemented modeling of memory updates [2 loops] (2) arithmetic instructions that cannot be handled by *yices* [4 loops] (3) insufficient invariant analysis [4 loops] (4) quantified invariants on array contents needed [6 loops in 2 programs].

The cBench benchmark was collected for research on program and compiler optimization. After removing code-duplicates it contains 1027 C source code files, 211.892 lines of code and a total number of 4302 loops. For 4090 loops our tool answered within a 1000 seconds timeout (3923 loops in less than 4 seconds). On 71 loops our tool exceeded the 1000 seconds timeout and 141 loops could not be analyzed because our current tool does not handle irreducible CFGs.

Our tool computed a bound for 75% of the 4090 analyzed loops in the cBench benchmark. On the non-trivial loops bound computation was successful in 65% of the cases (1181 of 1902 loops). For the class of inner loops (e.g. program location $l_2$ in Example 1) we were able to compute a bound for 65% (830 of 1345) of the loops. This class of loops is especially interesting for evaluating the precision of the automatically computed bounds in the presence of outer loops. A manual sample of around 100 loops in this class showed that the bounds by our tool were precise.

We evaluated our transitive hull algorithm on the class of loops for which an inner loop had to be summarized in order to compute an iteration bound. The bound computation was successful in 56% of these cases (578 of 1102). The relatively low success ratio of 56% is caused by limits of our implementation of the transitive hull algorithm that currently does not support invariants involving values of memory locations.

In 992 of the total 1017 failure cases we failed to compute a bound because we could not find a local ranking function that proves the termination of a single transition. Recall that such local ranking functions are used as norms in our bound analysis. A manual analysis revealed that the reasons for failure were: (1) missing implementation features like pointer calculations and memory updates (2) insufficient invariant analysis (3) some loops were not meant to terminate, e.g. input loops (4) complex invariants like quantified invariants on the content of arrays needed. None of these reasons reveals a general limitation of our method. All but reason (4) can be solved by systematic engineering work. In the 25 remaining cases our tool computed a bound for each transition but was not able to compose an overall bound.

**Acknowledgement.** We would like to thank the anonymous reviewers for their insightful comments.

# References

1. http://ctuning.org/wiki/index.php/CTools:CBench.
2. J. H. 0002, K. Aehlig, and M. Hofmann. Multivariate amortized resource analysis. In *POPL*, pages 357–370, 2011.
3. A. M. Ben-Amram. A complexity tradeoff in ranking-function termination proofs. *Acta Inf.*, 46(1):57–72, 2009.
4. A. M. Ben-Amram. Monotonicity constraints for termination in the integer domain. Technical report, 2011.
5. J. Berdine, A. Chawdhary, B. Cook, D. Distefano, and P. W. O'Hearn. Variance analyses from invariance analyses. In *POPL*, pages 211–224, 2007.
6. D. Beyer, A. Cimatti, A. Griggio, M. E. Keremoglu, and R. Sebastiani. Software model checking via large-block encoding. In *FMCAD*, pages 25–32, 2009.
7. C. Colby and P. Lee. Trace-based program analysis. In *POPL*, pages 195–207, 1996.
8. B. Cook, A. Podelski, and A. Rybalchenko. Termination proofs for systems code. In *PLDI*, pages 415–426, 2006.
9. B. Dutertre and L. de Moura. The yices smt solver. Technical report, 2006.
10. S. Goldsmith, A. Aiken, and D. S. Wilkerson. Measuring empirical computational complexity. In *ESEC/SIGSOFT FSE*, pages 395–404, 2007.
11. D. Gopan and T. W. Reps. Lookahead widening. In *CAV*, 2006.
12. B. S. Gulavani and S. Gulwani. A numerical abstract domain based on expression abstraction and max operator with application in timing analysis. In *CAV*, pages 370–384, 2008.
13. S. Gulwani. Speed: Symbolic complexity bound analysis. In *CAV*, pages 51–62, 2009.
14. S. Gulwani, S. Jain, and E. Koskinen. Control-flow refinement and progress invariants for bound analysis. In *PLDI*, pages 375–385, 2009.
15. S. Gulwani, K. K. Mehra, and T. M. Chilimbi. Speed: precise and efficient static estimation of program computational complexity. In *POPL*, pages 127–139, 2009.
16. S. Gulwani and A. Tiwari. Computing procedure summaries for interprocedural analysis. In *ESOP*, pages 253–267, 2007.
17. S. Gulwani and F. Zuleger. The reachability-bound problem. In *PLDI*, pages 292–304, 2010.
18. M. Heizmann, N. D. Jones, and A. Podelski. Size-change termination and transition invariants. In *SAS*, pages 22–50, 2010.
19. S. Jost, K. Hammond, H.-W. Loidl, and M. Hofmann. Static determination of quantitative resource usage for higher-order programs. In *POPL*, pages 223–236, 2010.
20. A. Krauss. Certified size-change termination. In *CADE*, pages 460–475, 2007.
21. D. Kroening, N. Sharygina, S. Tonetta, A. Tsitovich, and C. M. Wintersteiger. Loop summarization using abstract transformers. In *ATVA*, pages 111–125, 2008.
22. D. Kroening, N. Sharygina, A. Tsitovich, and C. M. Wintersteiger. Termination analysis with compositional transition invariants. In *CAV*, pages 89–103, 2010.
23. C. Lattner and V. Adve. Llvm: A compilation framework for lifelong program analysis & transformation. In *CGO '04: Proceedings of the international symposium on Code generation and optimization*, page 75, Washington, DC, USA, 2004. IEEE Computer Society.
24. C. S. Lee, N. D. Jones, and A. M. Ben-Amram. The size-change principle for program termination. In *POPL*, pages 81–92, 2001.

25. P. Manolios and D. Vroon. Termination analysis with calling context graphs. In *CAV*, pages 401–414, 2006.
26. D. Monniaux. Automatic modular abstractions for linear constraints. In *POPL*, pages 140–151, 2009.
27. A. Podelski and A. Rybalchenko. Transition invariants. In *LICS*, pages 32–41, 2004.
28. A. Podelski and A. Rybalchenko. Transition predicate abstraction and fair termination. In *POPL*, pages 132–144, 2005.
29. C. Popeea and W.-N. Chin. Inferring disjunctive postconditions. In *ASIAN*, pages 331–345, 2006.
30. A. Tsitovich, N. Sharygina, C. M. Wintersteiger, and D. Kroening. Loop summarization and termination analysis. In *TACAS*, pages 81–95, 2011.
31. R. Wilhelm, J. Engblom, A. Ermedahl, N. Holsti, S. Thesing, D. B. Whalley, G. Bernat, C. Ferdinand, R. Heckmann, T. Mitra, F. Mueller, I. Puaut, P. P. Puschner, J. Staschulat, and P. Stenström. The worst-case execution-time problem - overview of methods and survey of tools. *ACM Trans. Embedded Comput. Syst.*, 7(3), 2008.

## A  Comparison between Blockwise and Pathwise SCA Analysis

Classical papers on termination analysis with SCA [24,3] do not discuss how one can obtain abstract programs. However, these papers assume that abstract programs are given as CFGs whose edges are labeled by SCRs. Therefore it is fair to say that classical SCA uses blockwise analysis. We sketch how this abstraction strategy differs from our approach: Classical SCA abstracts program in one step, and then analyzes the abstracted programs by a single transitive hull computation. In contrast, our pathwise analysis abstracts programs in multiple steps and computes transitive hulls at multiple times during the analysis. Our approach is a generalization of classical SCA and strictly more precise.

We illustrate the difference between our pathwise analysis and classical SCA in the following. Let $P$ be the program in Example 1. As described in Section B.1 we obtain the transition system $\texttt{TransSys}(P, l_1) = \{n - i - 1 > 0 \land n' - i' < n - i \land j' > 0, n - i > 0 \land n' - i' = n - i - 1 \land j' = 0\}$ for $P|_{l_1}$ by pathwise analysis. Note that $\texttt{TransSys}(P, l_1)$ establishes that the variable $i$ increases at every loop iteration and that $i < n$ is an invariant at $l_1$. $\texttt{TransSys}(P, l_1)$ is precise enough so that our Algorithm 2 can further size-change abstract it and compute a bound from the abstraction.

The blockwise analysis in classical SCA begins with abstracting $P$. Because of the inner loop at location $l_2$ of program $P$, each transition $\rho_1, \rho_2, \rho_3, \rho_4$ constitutes a program block and needs to be abstracted separately. We get the SCRs $\alpha(\rho_1) = n - i > 0 \land n' - i' < n - i \land j' \geq 0, \alpha(\rho_2) = n - i > 0 \land n' - i' < n - i \land j < j', \alpha(\rho_3) = j > 0 \land n' - i' > n - i, \alpha(\rho_4) = j \leq 0 \land n' - i' = n - i$. The termination analysis with classical SCA computes the transitive hull of these SCRs along the control flow edges of program $P$. In particular classical SCA computes the SCR $\alpha(\rho_1) \circ \alpha(\rho_2) \circ \alpha(\rho_3) = n - i > 0 \land j' > 0$ for the path $l_1 \xrightarrow{\rho_1} l_2 \xrightarrow{\rho_2} l_2 \xrightarrow{\rho_3} l_1$. Note that classical SCA cannot establish that $n - i$ increases every time this path is taken (the concatenation of $n' - i' < n - i$ and $n' - i' > n - i$ in $\alpha(\rho_2)$ and $\alpha(\rho_3)$ loses all information on $n - i$) and therefore cannot prove the termination of $P$.

## B  Examples

### B.1  Example application of Algorithm 1

Let $P$ be the program of Example 1. We want to compute the transition system for $P|_{l_1}$ and hence call $\text{TransSys}(P, l_1)$. In the first `foreach`-loop Algorithm 1 calls itself recursively on the nested loop $loop = (\{l_2\}, \{l_2 \xrightarrow{\rho_2} l_2\})$ with header $l_2$. In the recursive call Algorithm 1 skips the first `foreach`-loop because $(loop, l_2)$ does not have nested loops. The second `foreach`-loop iterates over all cycle-free paths of $\text{paths}(loop, l_2)$. There is only one such a path $\pi = l_2 \xrightarrow{\rho_2} l_2$. Algorithm 1 computes $\mathcal{T}_\pi = \{i < n \wedge i' = i + 1 \wedge j' = j + 1 \wedge n' = n\}$ and returns $\mathcal{T}_\pi$ as the transition system $\text{TransSys}(loop, l_2)$. After the return of the recursive call $\mathcal{T} = \text{TransSys}(loop, l_2)$ Algorithm 1 size-change abstracts $\mathcal{T}$ by $\alpha(\mathcal{T}) = \{n - i > 0 \wedge n' - i' < n - i \wedge j < j'\}$, computes the reflexive transitive hull in the abstract $\alpha(\mathcal{T})^* = \{n' - i' = n - i \wedge j' = j, n - i > 0 \wedge n' - i' < n - i \wedge j' > j\}$ and stores $\gamma(\alpha(\mathcal{T})^*)$ in $\text{summary}[l_2]$. As there is no other nested loop the first `foreach`-loop is finished. The second `foreach`-loop iterates over all cycle-free paths of $\text{paths}(P, l_1)$. There are two such paths $\pi_1, \pi_2$ as explained in Section 2. Algorithm 1 computes $\mathcal{T}_{\pi_1} = \{\rho_1\} \circ \text{summary}[l_2] \circ \{\rho_3\} = \{n - i > 0 \wedge i_1 = i + 1 \wedge j_1 = 0 \wedge n' - i' = n - i \wedge j' = j_1 \wedge j' > 0, i_1 = i + 1 \wedge j_1 = 0 \wedge n - i_1 > 0 \wedge n' - i' < n - i \wedge j' > j_1 \wedge j' > 0\} = \{false, n - i - 1 > 0 \wedge n' - i' < n - i \wedge j' > 0\}$, $\mathcal{T}_{\pi_2} = \{\rho_1\} \circ \text{summary}[l_2] \circ \{\rho_4\} = \{n - i > 0 \wedge i_1 = i + 1 \wedge j_1 = 0 \wedge n' - i' = n - i_1 \wedge j' = j_1 \wedge j' = 0, i_1 = i + 1 \wedge j_1 = 0 \wedge n - i_1 > 0 \wedge n' - i' > n - i_1 \wedge j' > j_1 \wedge j' = 0\} = \{n - i > 0 \wedge n' - i' = n - i - 1 \wedge j' = 0, false\}$ and returns $\mathcal{T}_{\pi_1} \cup \mathcal{T}_{\pi_2} = \{n - i - 1 > 0 \wedge n' - i' < n - i \wedge j' > 0, n - i > 0 \wedge n' - i' = n - i - 1 \wedge j' = 0\}$ as transition system for $P|_{l_1}$.

### B.2  Example application of Algorithm 2

The CFG in Figure 2 has 5 SCCs: $(l_1)$, $(l_2)$, $(l_3)$, $(l_4)$, $(l_5, l_6)$. Algorithm 3 computes the following bounds on these SCCs: $b_{l_1} = \max(255 - s, 0)$, $b_{l_2} = \max(s, 0)$, $b_{l_3} = 1$, $b_{l_4} = 1$, $b_{l_5, l_6} = l = \log c$. Algorithm 2 composes these bounds as follows: $\max(u(b_{l_1}), u(b_{l_2})) + \max(u(b_{l_3}), u(b_{l_4})) + u(b_{l_5, l_6})$, where $u$ denotes an upper bound on the value of the given expression computed by an invariant analysis. Assuming that the invariant analysis provides $u(b_{l_1}) = 255$, $u(b_{l_2}) = s$, $u(b_{l_3}) = 1$, $u(b_{l_4}) = 1$ and $u(b_{l_5, l_6}) = 2$, we obtain the precise bound $\max(255, s) + 3$.

### B.3  Flagship Example of [14]

In this subsection we apply our analysis to the flagship example of a recent publication [14] on the bound problem (Example 4 below). On this example the authors of [14] motivate control-flow refinement for bound analysis. Their algorithm relies on a sophisticated interplay between control-flow refinement and abstract interpretation. We show that our simpler technique can also handle the example.

*Example 4.* `void cyclic(int id, int maxId) {`
`assume(0 <= id <= maxId);`
`int tmp := id+1;`
`while(tmp !=id && nondet()){`
`    if (tmp <= maxId) tmp := tmp + 1;`
`    else tmp := 0; } }`

We assume an invariant analysis (e.g. octagon analysis) provides the invariant $0 \leq id \leq maxid \wedge tmp \geq 0$ at the header of the `while`-loop. The `while`-loop has four paths because we consider the inequality `tmp != id` as the disjunction $tmp < id \vee tmp > id$. Algorithm 1 gives us the transition system $\mathcal{T} = \{\rho_1, \rho_2, \rho_3\}$ (there is no fourth transition because $0 \leq id \leq maxid \wedge tmp < id \wedge tmp > maxid$ is unsatisfiable), where

$\rho_1 \equiv 0 \leq id \leq maxid \wedge id < tmp \leq maxid \wedge tmp' = tmp + 1,$
$\rho_2 \equiv 0 \leq id \leq maxid \wedge tmp > maxid \wedge tmp' = 0,$
$\rho_3 \equiv 0 \leq id \leq maxid \wedge 0 \leq tmp < id \wedge tmp' = tmp + 1.$

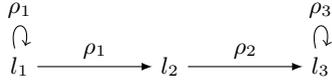

**Fig. 4.** Contextualization of Example 4

Contextualization gives us the CFG depicted in Figure 4, which precisely reflects the different phases of the loop.

The control flow graph given in Figure 4 has 3 SCCs: $(l_1), (l_2), (l_3)$. Algorithm 3 computes the following bounds on these SCCs: $b_{l_1} = \max(maxId - tmp, 0)$, $b_{l_2} = 1$ and $b_{l_3} = \max(id - tmp, 0)$. Algorithm 2 composes these bounds as follows: $u(b_{l_1}) + u(b_{l_2}) + u(b_{l_3})$, where $u$ denotes an upper bound on the value of the given expression. Assuming that the invariant analysis provides $u(b_{l_1}) = maxId - id$, $u(b_{l_2}) = 1$, $u(b_{l_3}) = id$, we obtain the precise bound $maxId + 1$.

## C Proof of Theorem 2

We prove a stronger statement than the one stated in Theorem 2: for every program $P = (L, E)$ and location $l \in L$ it holds that `TransSys`$(P, l)$ is a transition system for $P|_l$ and that for every inner loop *loop* of $P$ w.r.t. $l$ with header *header* the transition set summary[*header*] is a transition invariant for $loop|_{header}$ (*).

The stronger statement has the advantage to be inductive, whereas the statement of Theorem 2 is not. Our proof of (*) proceeds by induction on the loop nesting structure of programs.

Let $P = (L, E)$ be a program and $l \in L$ be some location. In the base case, there are no inner loops of $P$ w.r.t. $l$. Let $\pi = l \xrightarrow{\rho_0} l_1 \xrightarrow{\rho_1} \cdots l_k \xrightarrow{\rho_k} l \in \mathtt{paths}(P, l)$ be a path with start and end location $l$. Because $P$ does not have inner loops w.r.t. $l$, $\pi$ is cycle free. Thus, no location $l_i$ is the header of an inner loop. Therefore we have $\mathtt{ITE}(\mathtt{IsHeader}(l_i), \mathsf{summary}[l_i], \{Id\}) = \{Id\}$ for all $i$. Hence,

$$
\begin{aligned}
\{\mathtt{rel}(\pi)\} &= \{\rho_0 \circ \rho_1 \circ \rho_2 \circ \cdots \circ \rho_k\} \\
&= \{\rho_0\} \circ \{Id\} \circ \{\rho_1\} \circ \{Id\} \circ \{\rho_2\} \circ \cdots \circ \{Id\} \circ \{\rho_k\} \\
&= \{\rho_0\} \circ \mathtt{ITE}(\mathtt{IsHeader}(l_1), \mathsf{summary}[l_1], \{Id\}) \circ \\
&\quad \{\rho_1\} \circ \mathtt{ITE}(\mathtt{IsHeader}(l_2), \mathsf{summary}[l_2], \{Id\}) \circ \{\rho_2\} \circ \cdots \circ \\
&\quad \mathtt{ITE}(\mathtt{IsHeader}(l_k), \mathsf{summary}[l_k], \{Id\}) \circ \{\rho_k\}
\end{aligned}
$$

Therefore,

$$
\bigcup \mathtt{TransSys}(P, l) = \bigcup_{\text{cycle-free path } \pi \in \mathtt{paths}(P, l)} \mathtt{rel}(\pi) = \bigcup_{\pi \in \mathtt{paths}(P, l)} \mathtt{rel}(\pi) = P|_l
$$

.
Thus $\mathtt{TransSys}(P, l)$ is a transition system for $P|_l$.

In the inductive case, there are nested loops of $P$ w.r.t. $l$. Let $loop$ be a nested loop of $P$ w.r.t. $l$ and let $header$ be its header.

We show that $\mathsf{summary}[header]$ is a transition invariant for $loop|_{header}$. By the induction hypothesis we have that $\mathcal{T} = \mathtt{TransSys}(loop, header)$ is a transition system for $loop|_{header}$. Because $\mathcal{T}$ is a transition system for $loop|_{header}$, we have that $\gamma(\alpha(\mathcal{T})^*)$ is a transition invariant for $loop|_{header}$ by the soundness of SCA as stated in Theorem 1. With $hull := \gamma(\alpha(\mathcal{T})^*)$ and $\mathsf{summary}[header] := hull$, $\mathsf{summary}[header]$ is a transition invariant for $loop|_{header}$.

By the induction hypothesis we further have that for all inner loops $loop'$ of $loop$ w.r.t $header$ with header $header'$ the transition set $\mathsf{summary}[header']$ is a transition invariant for $loop'|_{header'}$.

Thus we have that for every inner loop $loop$ of $P$ w.r.t. $l$ with header $header$ the transition set $\mathsf{summary}[header]$ is a transition invariant for $loop|_{header}$.

It remains to show that $\mathtt{TransSys}(P, l)$ is a transition system for $P|_l$. It suffices to show that we have $\mathtt{rel}(\pi) \subseteq \bigcup \mathtt{TransSys}(P, l)$ for every path $\pi \in \mathtt{paths}(P, l)$. Let $\pi = l \xrightarrow{\rho_0} l_1 \xrightarrow{\rho_1} \cdots l_k \xrightarrow{\rho_k} l \in \mathtt{paths}(P, l)$ be a path of $P$ with start and end location $l$. In the following we iteratively remove iterations through inner loops from $\pi$ to obtain a cycle-free path. Let $i_1$ be the first index such that $l_{i_1}$ appears multiple times in $\pi$. Let $loop_1$ be the innermost loop of $P$ w.r.t. $l$ that contains $l_{i_1}$. Because $P$ w.r.t. $l$ is reducible, there is a unique loop header *header* of $loop_1$. Because *header* is a dominator for $l_{i_1}$ every path from $l$ to $l_{i_1}$ must visit *header* before. Because $loop_1$ is the innermost loop that contains $l_{i_1}$, every path of $P$ that starts and ends in $l_{i_1}$ must visit *header*. Therefore $\pi$ also visits *header* multiple times. As $l_{i_1}$ is the first location visited multiple times we must have $l_{i_1} = header$. Let $j_1$ be the last index such that $l_{j_1} = l_{i_1}$. We denote by $q_1 = \pi[i_1, j_1] = l_{i_1} \xrightarrow{\rho_{i_1}} l_{i_1+1} \xrightarrow{\rho_{i_1+1}} \cdots l_{j_1-1} \xrightarrow{\rho_{j_1}} l_{j_1}$ the subpath of $\pi$ from index $i_1$ to index $j_1$. We have that $q_j \in \texttt{paths}(loop_1, l_{i_1})$ is some iteration through the inner loop $loop_1$ with header $l_{i_1}$. Let $\pi_1$ be the result of deleting the subpath from index $i_1 + 1$ to index $j_1$ of $\pi$. $\pi_1$ is a path because $l_{i_1} = l_{j_1}$. Note that $\pi_1$ does not contain $l_{i_1}$ multiple times any more, but does contain $l_{i_1}$ exactly once. We iterate this approach to derive indices $i_2, j_2, i_3, j_3, \ldots, i_m, j_m$ and paths $q_2, \pi_2, q_3, \pi_3, \ldots, q_m, \pi_m$ until $\pi_m$ does not contain a location that appears multiple times. By induction assumption we have that $\textsf{summary}[l_{i_j}]$ is a transition invariant for $loop_j|_{l_{i_j}}$ for all $1 \leq j \leq m$. Thus $\texttt{rel}(q_j) \subseteq loop_j|_{l_{i_j}}^* \subseteq \bigcup \textsf{summary}[l_{i_j}]$ for all $1 \leq j \leq m$. This gives us

$$\begin{aligned}
\texttt{rel}(\pi) &= \rho_0 \circ \rho_1 \circ \cdots \circ \rho_k \\
&= \rho_0 \circ \rho_1 \circ \cdots \circ \rho_{i_1-1} \circ \rho_{i_1} \circ \cdots \circ \rho_{j_1} \circ \rho_{j_1+1} \circ \cdots \\
&\quad \circ \rho_{i_m-1} \circ \rho_{i_m} \circ \cdots \circ \rho_{j_m} \circ \rho_{j_m+1} \circ \cdots \circ \rho_k \\
&= \rho_0 \circ \rho_1 \circ \cdots \circ \rho_{i_1-1} \circ \texttt{rel}(q_1) \circ \rho_{j_1+1} \circ \cdots \\
&\quad \circ \rho_{i_m-1} \circ \texttt{rel}(q_m) \circ \rho_{j_m+1} \circ \cdots \circ \rho_k \\
&\subseteq \bigcup (\{\rho_0\} \circ \{\rho_1\} \circ \cdots \circ \{\rho_{i_1-1}\} \circ \textsf{summary}[l_{i_1}] \circ \{\rho_{j_1+1}\} \circ \cdots \\
&\quad \circ \{\rho_{i_m-1}\} \circ \textsf{summary}[l_{i_m}] \circ \{\rho_{j_m+1}\} \circ \cdots \circ \{\rho_k\}) \\
&= \bigcup (\{\rho_0\} \circ \texttt{ITE}(\texttt{IsHeader}(l_1), \textsf{summary}[l_1], \{Id\}) \circ \{\rho_1\} \\
&\quad \circ \texttt{ITE}(\texttt{IsHeader}(l_2), \textsf{summary}[l_2], \{Id\}) \circ \{\rho_2\} \circ \cdots \\
&\quad \circ \texttt{ITE}(\texttt{IsHeader}(l_k), \textsf{summary}[l_k], \{Id\}) \circ \{\rho_k\}) \\
&= \bigcup \texttt{TransSys}(P, l).
\end{aligned}$$

This concludes the proof of (*).